\documentclass{preprint}

\usepackage{epsfig}

\usepackage{amssymb}

\newcommand{\rs}[1]{{\tt DRSP}{(#1)}}

\begin{document}

\date{31 October 2002}

\begin{frontmatter}



\title{Linear-Time Algorithms for Computing Maximum-Density Sequence Segments
with Bioinformatics Applications%
\thanksref{label1}%
} 

\thanks[label1]{A significant portion of these results appeared under the title, 
``Fast Algorithms for Finding Maximum-Density Segments of a
Sequence with Applications to Bioinformatics,''
in {\em Proceedings of the Second Workshop on
Algorithms in Bioinformatics (WABI)}, volume~2452 of Lecture Notes in
Computer Science (Springer-Verlag, Berlin),
R.~Guig\'{o} and D.~Gusfield editors, 2002, pp.~157--171.
}

\author{Michael H.~Goldwasser
}
\address{
Department of Computer Science\\
Loyola University Chicago\\
6525 N.~Sheridan Rd.\\
Chicago, IL 60626.\\
Email: {mhg@cs.luc.edu}\\
URL: {www.cs.luc.edu/\~{ }mhg}
}

\author{Ming-Yang Kao\thanksref{ming}}
\address{
Department of Computer Science\\
Northwestern University\\
Evanston, IL 60201.\\
Email: {kao@cs.northwestern.edu}\\
URL:  {www.cs.northwestern.edu/\~{ }kao}
}
\thanks[ming]{Supported in part by NSF grant EIA-0112934.}

\author{Hsueh-I Lu\thanksref{hsuehI}}
\address{
Institute of Information Science\\ 
Academia Sinica\\ 
128 Academia Road, Section 2\\
Taipei 115, Taiwan.\\
Email: {hil@iis.sinica.edu.tw}\\
URL:  {www.iis.sinica.edu.tw/\~{ }hil}
}
\thanks[hsuehI]{Supported in part by NSC grant
NSC-90-2218-E-001-005.}

\end{frontmatter}

\newpage
\begin{frontmatter}

\begin{abstract}
We study an abstract optimization problem arising from biomolecular
sequence analysis.  For a sequence $A$ of pairs $(a_i,w_i)$ for $i = 1,
\ldots, n$ and $w_i>0$, a {\em segment} $A(i,j)$ is a consecutive 
subsequence of $A$ starting with index $i$ and ending with index $j$.
The {\em width} of $A(i,j)$ is $w(i,j) = \sum_{i \leq k \leq j} w_k$,
and the {\em density} is \((\sum_{i\leq k \leq j} a_k)/ w(i,j).\) The
{\it maximum-density segment} problem takes $A$ and two values $L$ and
$U$ as input and asks for a segment of $A$ with the largest possible
density among those of width at least $L$ and at most $U$.  When $U$
is unbounded, we provide a relatively simple, $O(n)$-time algorithm,
improving upon the $O(n \log L)$-time algorithm by Lin, Jiang and
Chao.  When both $L$ and $U$ are specified, there are no previous
nontrivial results.  We solve the problem in $O(n)$ time if $w_i=1$
for all $i$, and more generally in $O(n+n\log(U-L+1))$ time when $w_i \geq
1$ for all $i$.

\end{abstract}


\end{frontmatter}


\section{Introduction}\label{sec:intro}
Non-uniformity of nucleotide composition within genomic sequences was
first revealed through thermal melting and gradient centrifugation%
~\cite{Inman:1966:DMP,Macaya:1976:AOE}.  The GC content of the DNA
sequences in all organisms varies from 25\% to 75\%. GC-ratios have
the greatest variations among bacteria's DNA sequences, while the
typical GC-ratios of mammalian genomes stay in 45-50\%.  The GC
content of human DNA varies widely throughout the genome, ranging
between 30\% and 60\%.  Despite intensive research effort in the past
two decades, the underlying causes of the observed heterogeneity
remain
contested~\cite{Barhardi:2000:IEG,Bernardi:1986:CCG,Charlesworth:1994:GRP,Eyre-Walker:1992:EBG,Eyre-Walker:1993:RMG,Filipski:1987:CBM,Francino:1999:IRM,Holmquist:1992:CBT,Sueoka:1988:DMP,Wolfe:1989:MRD}.
Researchers~\cite{Nekrutenko:2000:ACH,Stojanovic:1999:CFM} observed that the
compositional heterogeneity is highly correlated to the GC content of
the genomic sequences.  Other investigations showed that gene
length~\cite{Duret:1995:SAV}, gene density~\cite{Zoubak:1996:GDH}, patterns of
codon usage~\cite{Sharp:1995:DSE}, distribution of different classes of
repetitive elements~\cite{Duret:1995:SAV,Soriano:1983:DIR}, number of
isochores~\cite{Barhardi:2000:IEG}, lengths of
isochores~\cite{Nekrutenko:2000:ACH}, and recombination rate within
chromosomes~\cite{Fullerton:2001:LRR} are all correlated with GC content.
More research exists related to GC-rich segments
\cite{Guldberg:1998:DMG,Henke:1997:BIP,Ikehara:1996:PON,Jin:1997:WIN,Murata:2001:TAF,Madsen:1997:ICE,Scotto:1993:GRD,Wang:2002:RFP,Wu:1999:IMD}.

Although GC-rich segments of DNA sequences are important in gene
recognition and comparative genomics, only a couple of algorithms for
identifying GC-rich segments appeared in the literature. A widely used
window-based approach is based upon the GC-content statistics of a
fixed-length
window~\cite{Fields:1990:GPT,Hardison:1991:SCA,Nekrutenko:2000:ACH,Rice:2000:EEM}. Due
to the fixed length of windows, these practically fast approaches are
likely to miss GC-rich segments that span more than one window.
Huang~\cite{Huang:1994:AIR} proposed an algorithm to accommodate
windows with variable lengths. Specifically, by assigning $-p$ points
to each AT-pair and $1-p$ points to each GC-pair, where $p$ is a
number with $0\leq p\leq 1$, Huang gave a linear-time algorithm for
computing a segment of length no less than $L$ whose score is
maximized. As observed by Huang, however, this approach tends to
output segments that are significantly longer than the given $L$.

In this paper, we study the following abstraction of the problem. Let
$A$ be a sequence of pairs $(a_i,w_i)$ for $i = 1, \ldots, n$ and $w_i
> 0$. A {\it segment} $A(i,j)$ is a consecutive subsequence of $A$
starting with index $i$ and ending with index $j$.  The {\em width} of
$A(i,j)$ is $w(i,j) = \sum_{i \leq k \leq j} w_k$, and the {\em
density} is \((\sum_{i\leq k \leq j} a_k)/ w(i,j).\) Let $L$ and $U$
be positive values with $L \leq U$.  The {\it maximum-density segment}
problem takes $A$, $L$, and $U$ as input and asks for a segment of $A$
with the largest possible density among those of width at least $L$
and at most $U$.  This generalizes a previously studied model, which
we term the {\em uniform} model, in which $w_i=1$ for all $i$.  All of
the previous work discussed in this section involves the uniform
model. We introduce the generalized model as it might be used to
compress a sequence $A$ of real numbers to reduce its sequence length
and thus its density analysis time in practice or theory.

In its most basic form, the sequence $A$ corresponds to the given DNA
sequence, where $a_i = 1$ if the corresponding nucleotide in the DNA
sequence is G or C; and $a_i = 0$ otherwise.  In the work of Huang,
sequence entries took on values of $p$ and $1-p$ for some real number
$0 \leq p \leq 1$.  More generally, we can look for regions where a
given set of patterns occur very often. In such applications, $a_i$
could be the relative frequency with which the corresponding DNA
character appears in the given patterns.  Further natural applications
of this problem can be designed for sophisticated sequence analyses
such as mismatch density~\cite{Sellers:1984:PRG}, ungapped local
alignments~\cite{Alexandrov:1998:SSU}, and annotated multiple sequence
alignments~\cite{Stojanovic:1999:CFM}.

Nekrutendo and Li~\cite{Nekrutenko:2000:ACH}, and Rice, Longden and
Bleasby~\cite{Rice:2000:EEM} employed algorithms for the case where $L
= U$. This case is trivially solvable in $O(n)$ time using a sliding
window of the appropriate length. More generally, when $L \neq U$,
this yields a trivial $O(n(U-L+1))$ algorithm.
Huang~\cite{Huang:1994:AIR} studied the case where $U = n$, i.e.,
there is effectively no upper bound on the width of the desired
maximum-density segments. He observed that an optimal segment exists
with width at most $2L-1$.  Therefore, this case is equivalent to the
case with $U=2L-1$ and thus can be solved in $O(nL)$ time.  Recently,
Lin, Jiang, and Chao~\cite{Lin:2002:EAL} gave an $O(n\log L)$-time
algorithm for this case based on the introduction of right-skew
partitions of a sequence.

In this paper, we present an $O(n)$-time algorithm which solves the
maximum-density segment problem in the absence of upper bound $U$.
When both lower and upper bounds, $L$ and $U$, are specified, we
provide an $O(n)$-time algorithm for the {\em uniform} case, and an
$O(n + n\log(U-L+1))$-time algorithm when $w_i \geq 1$ for all $i$.
Our results exploit the structure of locally optimal
segments to improve upon the $O(n \log L)$-time algorithm of Lin,
Jiang, and Chao~\cite{Lin:2002:EAL} and to extend the results to
arbitrary values of $U$.
The remainder of this paper is organized as follows.
Section~\ref{sec:notation} introduces some notation and definitions.
In Section~\ref{sec:skew}, we carefully review the previous work of
Lin, Jiang and Chao, in which they introduce the
concept of right-skew partitions. Our main results are presented
in Section~\ref{sec:results}.

Other related works include algorithms for the problem of computing a
segment $\langle a_i, \ldots a_j \rangle$ with a maximum sum
$a_i+\cdots+a_j$ as opposed to a maximum density.
Bentley~\cite{Bentley:1986:PP} gave an $O(n)$-time algorithm for the
case where $L=0$ and $U = n$.  Within the same linear time complexity,
Huang~\cite{Huang:1994:AIR} solved the case with arbitrary $L$ yet
unbounded $U$.  More recently, Lin, Jiang, and
Chao~\cite{Lin:2002:EAL} solved the case with arbitrary $L$ and $U$.

\section{Notation and Preliminaries}
\label{sec:notation}
We consider $A$ to be a sequence of $n$ objects, where each object is
represented by a pair of two real numbers $(a_i,w_i)$ for
$i = 1,\ldots,n$ and $w_i > 0$.
For $i \leq j$, we let $A(i,j)$ denote that segment of $A$ which
begins at index $i$ and ends with index $j$.
We let $w(i,j)$ denote the {\em width} of $A(i,j)$, defined as
$w(i,j) = \sum_{i \leq k \leq j} w_k$.
We let $\mu(i,j)$ denote the {\em density} of $A(i,j)$, defined as
$$
\mu(i,j) = \left(\sum_{i \leq k \leq j} a_k\right)/ w(i,j).$$
We note that the prefix sums of the input sequence can be
precomputed in $O(n)$ time.  With these, the values of $w(i,j)$ and
$\mu(i,j)$ can be computed in $O(1)$ time for any $(i,j)$ using the
following formulas,
\begin{eqnarray*}
w(i,j) \,\,\,\, & = & \,\,\,\,   \sum_{1\leq k \leq j} w_k - \sum_{1 \leq k \leq i-1} w_k,\\
\mu(i,j) \,\,\,\, & = & \,\,\,\, \left( \sum_{1\leq k \leq j} a_k -
\sum_{1 \leq k \leq i-1} a_k \right)/ w(i,j).
\end{eqnarray*}
The maximum-density segment problem is to find a segment $A(i,j)$ of
maximum density, subject to $L \leq w(i,j) \leq U$.  Without loss of
generality, we assume that $w_i \leq U$ for all $i$, as items with
larger width could not be used in a solution.  If $w_i=1$ for all $i$,
we denote this as the {\em uniform} model.

For a given index $i$, we introduce the notation $L_i$ for
the minimum index such that $w(i,L_i) \geq L$ if such an index exists,
and we let $U_i$ denote the maximum index such that $U_i \geq i$ and
$w(i,U_i) \leq U$.
A direct consequence of these definitions is that segment $A(i,j)$ has
width satisfying $L \leq w(i,j) \leq U$ if and only if $L_i$ is
well-defined and $L_i \leq j \leq U_i$.

In the uniform model, the set of all such values is easily calculated
in $O(n)$ time, as $L_i = i+L-1$ for $i \leq n-L+1$ and $U_i =
\min(i+U-1,n)$.
In general, the full set of $L_i$ and $U_i$ values can be precomputed in $O(n)$
time by a simple sweep-line technique.  The precomputation of the
$U_i$ values is shown in Figure~\ref{fig:Ui}; a similar technique can
be used for computing $L_i$ values.  It is not difficult to
verify the correctness and efficiency of these computations.

\begin{figure}[t]
\hrule
\begin{tabbing}
x\=xxx\=xxx\=xxx\=xxx\=xxx\=xxx\=\kill
\\
\>1\>$j \leftarrow n$\\
\>2\>{\bf for} $i \leftarrow n$ {\bf downto} $1$ {\bf do}\\
\>3\>\>{\bf while} $(w(i,j)>U)$ {\bf do}\\
\>4\>\>\>$j \leftarrow j-1$\\
\>5\>\>{\bf end while}\\
\>6\>\>$U_i \leftarrow j$\\
\>7\>{\bf end for}
\end{tabbing}
\hrule
\caption{Algorithm for precomputing $U_i$ for all $i$.}
\label{fig:Ui}
\end{figure}

\section{Right-Skew Segments}
\label{sec:skew}
For the uniform model,
Lin, Jiang and Chao~\cite{Lin:2002:EAL} define segment $A(i,k)$ to be
{\em right-skew} if and only if $\mu(i,j) \leq \mu(j+1,k)$ for all $i
\leq j < k$.  They define a partition of a sequence $A$ into segments $A_1A_2 \ldots A_m$
to be a {\em decreasingly right-skew partition} if it is the case that
each $A_i$ is right-skew, and that $\mu(A_x) > \mu(A_y)$ for any $x<y$.
The prove the following Lemma.
\begin{lem}
\label{lem:unique}
Every sequence $A$ has a unique decreasingly right-skew partition.
\end{lem}
We denote this unique partition as $\rs{A}$.
Within the proof of the above lemma, the authors implicitly
demonstrate the following fact.

\begin{lem}
\label{lem:decompLeft}
If segment $A(x,y)$ is not right-skew, then $\rs{A(x,y)}$ is precisely
equal to the union of $A(x,k)$ and $\rs{A(k+1,y)}$ where
$A(x,k)$ is the longest possible right-skew segement begining with
index $x$.
\end{lem}
Because of this structural property, the decreasingly right-skew
partitions of all suffixes of $A(1,n)$ can be simultaneously
represented by keeping a {\em right-skew pointer}, $p[i]$, for each $1
\leq i \leq n$.  The pointer is such that $A(i,p[i])$ is the
first right-skew segment of $\rs{A(i,n)}$.  They implicitly use
dynamic programming to construct all such right-skew pointers in
$O(n)$ time.

In order to find a maximum-density segment of width at least $L$, they
proceed by independently searching for the ``good partner'' of each
index $i$. The good partner of $i$ is the index $i'$ that maximizes
$\mu(i,i')$ while satisfying $w(i,i') \geq L$.  In order to find each
good partner, they make use of versions of the following three lemmas.

\begin{lem}[Atomic]
\label{lem:atomic}
Let $B$, $C$ and $D$ be sequences with $\mu(B)
\leq \mu(C) \leq \mu(D)$.  Then $\mu(BC) \leq \mu(BCD)$.
\end{lem}

\begin{lem}[Bitonic]
\label{lem:bitonic}
Let $B$ be a sequence and let $\rs{C} = C_1C_2\cdots C_m$ for
sequence $C$ which immediately follows $B$. Let $k$ be the greatest
index $i \in [0,m]$ that maximizes $\mu(BC_1C_2\cdots C_i)$.
Then $\mu(BC_1C_2\cdots C_i)$ $ > \mu(BC_1C_2\cdots C_{i+1})$ if and
only if $i \geq k$.
\end{lem}

\begin{lem}
\label{lem:2L}
Given a sequence $B$, let $C$ denote the shortest segment of $B$
realizing the maximum density for those segments of width at least
$L$.  Then the width of $C$ is at most $2L-1$.
\end{lem}

Without any upper bound on the desired segment length, the consequence
of these lemmas is an $O(\log L)$-time algorithm for finding a good
partner for arbitrary index $i$. Since only segments of width $L$ or
greater are of interest, the segment $A(i,L_i)$ must be included.  If
considering the possible inclusion of further elements,
Lemma~\ref{lem:atomic} assures that if part of a right-skew segment
increases the density, including that entire segment is just as
helpful (in the application of that lemma $C$ represents part of a
right-skew segment $CD$).  Therefore, the good partner for $i$ must be
$L_i$ or else the right endpoint of one of the right-skew segments
from $\rs{A(L_i+1,n)}$.  Lemma~\ref{lem:bitonic} shows that the
inclusion of each successive right-skew segment leads to a bitonic
sequence of densities, thus binary search can be used to locate the
good partner.  Finally, Lemma~\ref{lem:2L} assures that at most $L$
right-skew segments need be considered for inclusion, and thus the
binary search for a given $i$ runs in $O(\log L)$ time. The result is
an $O(n \log L)$-time algorithm for arbitrary $L$, with $U=n$.

Though presented in terms of the uniform model, the definition of a
right-skew segment involves only the densities of segments and so it
applies equally to our more general model.
Lemmas~{\ref{lem:unique}--\ref{lem:bitonic}} remain valid in the
general model.  A variant of Lemma~\ref{lem:2L} can be achieved with
the additional restriction that $w_i \geq 1$ for all $i$, and thus
their $O(n \log L)$-time algorithm applies subject to this additional
restriction.

\section{Improved Algorithms}
\label{sec:results}

Our techniques are built upon the use of decreasingly right-skew
partitions, as reviewed in Section~\ref{sec:skew}. Our improvements
are based upon the following observation.  An exact good partner for
an index $i$ need not be found if it can be determined that such a
partner would result in density no greater than that of a segment
already considered. This observation allows us to use a sweep-line
technique to replace the $O(\log L)$-time binary searches used by Lin,
Jiang and Chao~\cite{Lin:2002:EAL} with sequential searches that run
with an {\em amortized} time of $O(1)$.
In particular, we make use of the following key lemma.

\begin{lem}
\label{lem:main}
For a given $j$, assume $A(j,j')$ is a maximum-density segment of
those starting with index $j$, having $L \leq w(j,j') \leq U$, and
ending with index in a given range $[x,y]$.  For a given $i<j$, assume
$A(i,i')$ is a maximum-density segment of those starting with index
$i$, having $L \leq w(i,i') \leq U$ and ending in range $[x,y]$.
If $i'>j'$, then $\mu(j,j') \geq \mu(i,i')$.
\end{lem}

\begin{figure}[t]
\centerline{\psfig{figure=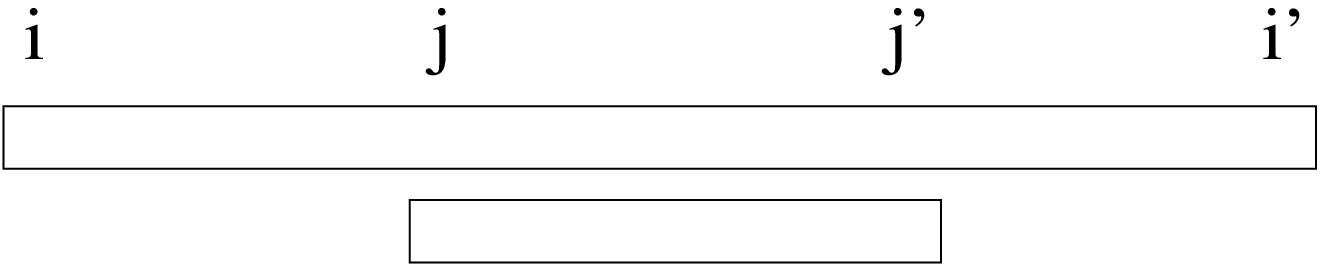,width=3.0in}}
\caption{Segments in proof of Lemma~\ref{lem:main}.}
\label{fig:lemma}
\end{figure}

\begin{pf}
A typical such configuration is shown in Figure~\ref{fig:lemma}.  By
assumption, both indices $i'$ and $j'$ lie within the range $[x,y]$.
Since $L \leq w(j,j') < w(j,i') < w(i,i') \leq U$, the optimality of
$A(j,j')$ guarantees that $\mu(j,j') \geq \mu(j,i')$.  This implies
that $\mu(j,j') \geq \mu(j,i') \geq \mu(j'+1,i')$.
Since $L \leq w(j,j') < w(i,j') < w(i,i') \leq U$, the
optimality of $A(i,i')$ guarantees that $\mu(i,i') \geq \mu(i,j')$, which
in turn implies $\mu(j'+1,i') \geq \mu(i,i') \geq \mu(i,j')$.
Combining these inequalities, $\mu(j,j') \geq \mu(j,i')
\geq \mu(j'+1,i') \geq \mu(i,i')$, thus proving the claim
that $\mu(j,j') \geq \mu(i,i')$.
\qed\end{pf}

Our high level approach is thus to find good partners for each left
endpoint $i$, considering those indices in decreasing order.  However,
rather than finding the true good partner for each $i$, our algorithm
considers only matching indices which are less than or equal to all
previously found good partners, in accordance with
Lemma~\ref{lem:main}.  In this way, as we sweep from right to left
over the left endpoints $i$, we also sweep from right to left over the
relevant matching indices. 

\subsection{Maximum-Density Segment with Width at Least $L$}
\label{sec:L}

In this section, we consider the problem of finding a segment with the
maximum possible density among those of width at least $L$.  We begin by
introducing a sweep-line data structure which helps manage the search
for good partners.

\subsubsection{A Sweep-Line Data Structure}
\label{sec:sweepL}

The data structure developed in this section is designed to answer
queries of the following type for a given range $[x,y]$, specified
upon initialization.  For left index $i$, the goal is to return a
matching right index $i'$ such that $\mu(i,i')$ is maximized, subject
to the constraints that $i' \in [x,y]$ and that $w(i,i') \geq L$. No
upper bound on the segment length is considered by this structure.

In order to achieve improved efficiency, the searches are limited in
the following two ways:
\begin{enumerate}
\item The structure can be used to find matches for many different
left indices, however such queries must be made in decreasing order.

\item When asked to find the match for a left index, the structure
only finds the true good partner in the case that the good partner has
index less than or equal to all previously returned indices.
\end{enumerate}

Our data structure augments the right-skew pointers for a given
interval with additional information used to speed up searches for
good partners.  The structure contains the following state
information, relative to given parameters $1 \leq x \leq y \leq n$:

\begin{itemize}

\item A (static) array, $p[k]$ for $x+1 \leq k \leq y$,  where
$A(k,p[k])$ is the {\em leftmost} segment of $\rs{A(k,y)}$.

\item A (static) sorted list, $S[k]$, for each $x+1 \leq k \leq y$,
containing all indices $j$ for which  $p[j]=k$.

\item Two indices $\ell$ and $u$ (for ``lower'' and
``upper''), whose values are non-increasing as the algorithm
progresses.

\item A variable, $b$ (for ``bridge''), which is maintained so
that $A(b,p[b])$ is the segment of $\rs{A(\ell,y)}$ which contains
index $u$.
\end{itemize}

\begin{figure}[t]
\hrule
\begin{tabbing}
x\=xxx\=xxx\=xxx\=xxx\=xxx\=xxx\=\kill
\\
\>\>{\bf procedure} {\tt InitializeL}$(x,y)$ 
\mbox{\hspace{0.75in}} {\em assumes $1 \leq x \leq y \leq n$}\\
\>1\>\>{\bf for} $i \leftarrow y$ {\bf downto} $x+1$ {\bf do}\\
\>2\>\>\>$S[i] \leftarrow \emptyset$\\
\>3\>\>\>$p[i] \leftarrow i$\\
\>4\>\>\>{\bf while} $\left((p[i]<y) \mbox{ and } (\mu(i,p[i])\leq \mu(p[i]+1,p[p[i]+1]))\right)$ {\bf do}\\
\>5\>\>\>\>$p[i] \leftarrow p[p[i]+1]$\\
\>6\>\>\>{\bf end while}\\
\>7\>\>\>Insert $i$ at beginning of $S[p[i]]$\\
\>8\>\>{\bf end for}\\
\>9\>\>$\ell \leftarrow y$; $u \leftarrow y$; $b \leftarrow y$
\end{tabbing}
\hrule
\caption{{\tt InitializeL} operation.}
\label{fig:initializeL}
\end{figure}

\begin{figure}[t]
\centerline{\psfig{figure=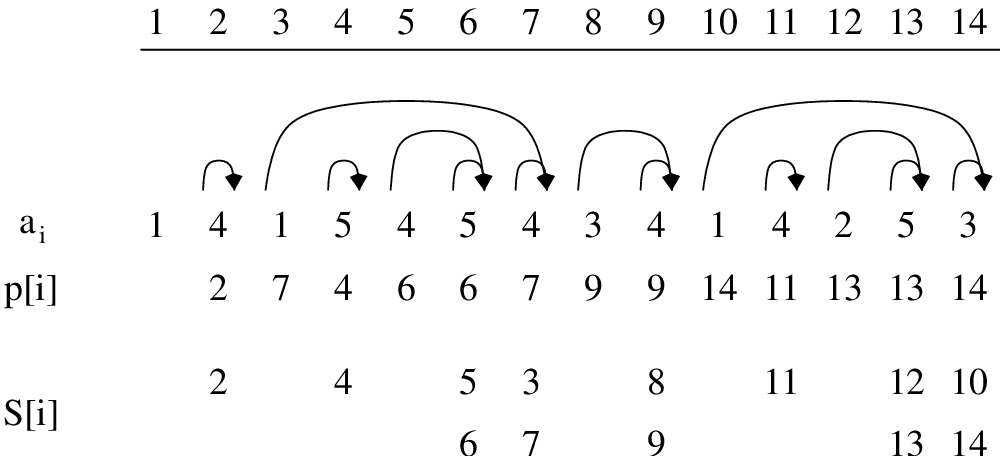,width=9cm}}
\caption{Example of data structure after {\tt InitializeL}$(1,14)$,
with $w_i = 1$ for all $i$.}
\label{fig:example}
\end{figure}

These data structures are initialized with procedure {\tt
InitializeL}$(x,y)$, given in Figure~\ref{fig:initializeL}. An example
of an initialized structure is given in Figure~\ref{fig:example}. 
Lines~1--8 of {\tt InitializeL} set the values $p[k]$ as was done in
the algorithm of Lin, Jiang and Chao~\cite{Lin:2002:EAL}.  Therefore,
we state the following fact, proven in that preceding paper.

\begin{lem}
\label{lem:p}
After a call to {\tt InitializeL}$(x,y)$, $p[k]$ is set for all $x+1
\leq k \leq y$ such that $A(k,p[k])$ is the leftmost segment of
$\rs{A(k,y)}$.
\end{lem}

We also prove the following nesting property of decreasingly
right-skew partitions.

\begin{lem}
\label{lem:nest-commonright}
Consider two segments $A(x_1,y)$ and $A(x_2,y)$ with a common right
endpoint.  Let $A(k,k')$ be a segment of $\rs{A(x_1,y)}$ and 
let $A(m,m')$ be a segment of $\rs{A(x_2,y)}$.  It cannot be the case
that $k < m \leq k' < m'$.
\end{lem}
\begin{pf}
If $A(k,k')$ is a segment of $\rs{A(x_1,y)}$, a repeated application
of Lemma~\ref{lem:decompLeft} assures that $A(k,k')$ is the leftmost
segment of $\rs{A(k,y)}$ and that $A(k,k')$ is the longest possible
right-skew segment of those starting with index $k$.

We assume for contradiction that $k < m \leq k' < m'$, and consider
the following three non-empty segments, $A(k,m-1)$, $A(m,k')$ and
$A(k'+1,m')$.  Since $A(k,k')$ is right-skew, it must be that
$\mu(k,m-1)
\leq \mu(m,k')$.  Since $A(m,m')$ is right-skew, it must be that
$\mu(m,k') \leq \mu(k'+1,m')$.  In this case, it must be that the
combined segment $A(k,m')$ is right-skew (this fact can be explicitly
proven by application of Lin, Jiang and
Chao's Lemma~4~\cite{Lin:2002:EAL}).  
Therefore the existence of right-skew segment $A(k,m')$ contradicts
the assumption that $A(k,k')$ is the longest right-skew segment
beginning with index $k$.
\qed\end{pf}

\begin{cor}
\label{cor:p}
There cannot exist indices $k$ and $m$ such that $k < m \leq p[k] < p[m]$.
\end{cor}
\begin{pf}
A direct result of Lemmas~\ref{lem:p}--\ref{lem:nest-commonright}.
\qed\end{pf}

We introduce the main query routine, {\tt FindMatchL}, given in
Figure~\ref{fig:findmatchL}.


\begin{figure}[t]
\hrule
\begin{tabbing}
x\=xxx\=xxx\=xxx\=xxx\=xxx\=xxx\=\kill
\\
\>\>{\bf procedure} {\tt FindMatchL}$(i)$\\
\>1\>\>{\bf while} $(\ell > 1+\max(x,L_i))$ {\bf do} \hspace{1.35in} // decrease $\ell$\\
\>2\>\>\>$\ell \leftarrow \ell-1$\\
\>3\>\>\>{\bf if} $(p[\ell] \geq u)$ {\bf then}\\
\>4\>\>\>\> $b \leftarrow \ell$\\
\>5\>\>\>{\bf end if}\\
\>6\>\>{\bf end while}\\
\>7\>\>{\bf while} $(u \geq \ell)$ and $(\mu(i,b-1) > \mu(i,p[b]))$
{\bf do} \hspace{0.25in} // bitonic search\\
\>8\>\>\>$u \leftarrow  b-1$\\
\>9\>\>\>{\bf if} $(u \geq \ell)$ {\bf then}\\
1\>0\>\>\>\>$b \leftarrow$ minimum $k \in S[u]$ such that $k \geq \ell$\\
1\>1\>\>\>{\bf end if}\\
1\>2\>\>{\bf end while}\\
1\>3\>\>{\bf return} $u$
\end{tabbing}
\hrule
\caption{{\tt FindMatchL}$(i)$ operation.}
\label{fig:findmatchL}
\end{figure}

\begin{lem}
\label{lem:bridge}
If $b$ is the minimum value satisfying $\ell \leq b \leq u \leq p[b]$,
then $A(b,p[b])$ is the segment of $\rs{A(\ell,y)}$ which contains index
$u$.
\end{lem}

\begin{pf}
By Lemma~\ref{lem:p}, $A(b,p[b])$ is the leftmost segment of
$\rs{A(b,y)}$, and as $b \leq u \leq p[b]$, $A(p,p[b])$ contains index
$u$.

By repeated application of Lemma~\ref{lem:decompLeft},
$\rs{A(\ell,y)}$ equals $A(\ell,p[\ell])$, $A(p[\ell]+1,
p[p[\ell]+1])$, and so on, until reaching right endpoint $y$.  We
claim that $A(b,p[b])$ must be part of that partition.  If not, there
must be some other $A(m,p[m])$ with $m < b \leq p[m]$.  By
Lemma~\ref{lem:nest-commonright}, it must be that $p[m] \geq p[b]$,
yet then we have $m < b \leq u \leq p[b] \leq p[m]$.  Such an $m$
violates the assumed minimality of $b$.
\qed\end{pf}

\begin{lem}
\label{lem:b}
Whenever line~7 of {\tt FindMatchL()} is evaluated, $b$ is the minimum
value satisfying $\ell \leq b \leq u \leq p[b]$, if such a value
exists.
\end{lem}
\begin{pf}
We show this by induction over time.  When initialized, $\ell = b = u
= p[b] = y$, and thus $b$ is the only satisfying value.  The only time
this invariant can be broken is when the value of $\ell$ or $u$
changes.  $\ell$ is changed only when decremented at line~2 of {\tt
FindMatchL}.  The only possible violation of the invariant would be if
the new index $\ell$ satisfies $\ell
\leq u \leq p[\ell]$.  This is exactly the condition handled by lines~3--4.

Secondly, $u$ is modified only at line~8 of {\tt FindMatchL}.
Immediately before this line is executed the invariant holds.  At this
point, we claim that $p[k] \leq b-1$ for any values of $k$ such that
$\ell \leq k < b$.  For $k<b$, Corollary~\ref{cor:p} implies that
either $p[k]<b$ or $p[k] \geq p[b]$.  If it were the case that $p[k]
\geq p[b] \geq u$ this would violate the minimality of $b$ assumed at
line~7. Therefore, it must be that $p[k] \leq b-1$ for all $\ell \leq k
\leq b-1$. As $u$ is reset to $b-1$, the only possible values for the
new bridge $b$ are those indices $k$ with $p[k]=u$, which is precisely
the set $S[b-1]$ considered at line~10 of {\tt FindMatchL}.
\qed\end{pf}

\begin{lem}
\label{lem:matchL}
Assume {\tt FindMatchL}$(i)$ is called with a value $i$ less than that
of all previous invocations and such that $L_i < y$.  Let $m_0$ be the
most recently returned value from {\tt FindMatchL}$()$ or $y$ if this
is the first such call. Let $A(i,m)$ be a maximum-density segment of
those starting with $i$, having width at least $L$, and ending with $m
\in [x,y]$.  Then {\tt FindMatchL}$(i)$ returns the value
$\min(m,m_0)$.
\end{lem}

\begin{pf}
All segments which start with $i$, having width at least $L$ and
ending with $m \in [x,y]$ must include interval $A(i,\max(x,L_i))$.
The loop starting at line~1 ensures that variable $\ell =
1+\max(x,L_i)$ upon the loop's exit.  As discussed in
Section~\ref{sec:skew}, the optimal such $m$ must either be $\ell-1$
or else among the right endpoints of $\rs{A(\ell,y)}$.

Since $u$ is only set within {\tt FindMatchL}, it must be that $u=m_0$
upon entering the procedure.  By Lemmas~\ref{lem:bridge}--\ref{lem:b},
$A(b,p[b])$ is the right-skew segment containing index $u$ in
$\rs{A(\ell,y)}$.  If $\mu(i,b-1) \leq
\mu(i,p[b])$, the good partner must have index at least $p[b] \geq u$,
by Lemma~\ref{lem:bitonic}.  In this case, the while loop is never
entered, and the procedure returns $m_0 =
\min(m,m_0)$.

In any other case, the true good partner for $i$ is less than or equal
to $m_0$, and this good partner is found by the while loop of
line~1, in accordance with Lemmas~\ref{lem:atomic}--\ref{lem:bitonic}.
\qed\end{pf}

\begin{lem}
\label{lem:matchL2}
If {\tt FindMatchL}$(i)$ returns value $i'$, it must be the case that for
some $j \geq i$, segment $A(j,i')$ is a maximum-density segment of those
starting with $j$, having width at least $L$, and ending in $[x,y]$.

\end{lem}
\begin{pf}
We prove this by induction over the number of previous calls to {\tt
FindMatchL}.  $i'=m$, as defined in the statement of
Lemma~\ref{lem:matchL}, then this claim is trivially true for $j=i$.
Otherwise, $i'$ is equal to the same value returned by the previous
call to {\tt FindMatchL}, and by induction, there is some $j \geq i$
such that segment $A(j,i')$ is such a maximum-density segment.
\qed\end{pf}

\begin{lem}
\label{lem:costL}
The data structure supports its operations with amortized running
times of $O(y-x+1)$ for {\tt InitializeL}$(x,y)$, and $O(1)$ for {\tt
FindMatchL}$(i)$.
\end{lem}
\begin{pf}
With the exception of lines~2,~7~and~9, the initialization procedure
is simply a restatement of the algorithm given by Lin, Jiang and
Chao~\cite{Lin:2002:EAL} for constructing the right-skew pointers.  An
$O(y-x+1)$-time worst-case bound was proven by those authors.

In analyzing the cost of {\tt FindMatchL} we note that variables $\ell$
and $u$ are initialized to value $y$ at line~9 of {\tt InitializeL}.
Variable $\ell$ is modified only when decremented at line~2 of {\tt
FindMatchL} and remains at least $x+1$ due to the condition at line~1.
Therefore, the loop of lines~1--6 executes at most $y-x+1$ times and
this cost can be amortized against the initialization cost.  Variable
$u$ is modified only at line~8. By Lemma~\ref{lem:b}, $x < \ell \leq b
\leq u \leq p[b]$, and so this line results in a strict decrease in the value
of $u$ yet $u$ remains at least $x$.  Therefore, the while loop of
lines~7--12 executes $O(y-x+1)$ times.  The only step within that loop
which cannot be bounded by $O(1)$ in the worst case is that of
line~10.  However, since each $k$ appears in list $S[u]$ for a
distinct value of $u$, the overall cost associated with line~10 is
bounded by $O(y-x+1)$.  Therefore the cost of this while loop can be
amortized as well against the initializaiton cost.  An $O(1)$
amortized cost per call can account for all remaining instructions
outside of the loops.
\qed\end{pf}

\subsubsection{An $O(n)$-time Algorithm}
\label{sec:algorithmL}

In Figure~\ref{fig:algorithmL}, we present a linear-time algorithm for
the maximum-density segment problem subject only to a lower bound of
$L$ on the segment width.  The algorithm makes use of the data
structure developed in Section~\ref{sec:sweepL}.

\begin{figure}[t]
\hrule
\begin{tabbing}
x\=xxx\=xxx\=xxx\=xxx\=xxx\=xxx\=\kill
\\
\>\>{\bf procedure} {\tt MaximumDensitySegmentL}$(A,L)$\\
\>1\>\>[calculate partial sums, $L_i$, as discussed
in Section~\ref{sec:notation}]\\
\>2\>\>{\bf call} {\tt InitializeL}$(1,n)$ to create data stucture\\
\>3\>\>$i_0 \leftarrow$ maximum index such that $L_{i_0}$ is well-defined\\
\>4\>\>{\bf for} $i \leftarrow i_0$ {\bf downto} $1$ {\bf do}\\
\>5\>\>\>{\bf if} $(L_i=y)$ {\bf then}  \hspace{1in} // only one feasible right index\\
\>6\>\>\>\>$g[i] \leftarrow y$\\
\>7\>\>\>{\bf else}\\
\>8\>\>\>\>$g[i] \leftarrow$ {\tt FindMatchL}$(i)$\\
\>9\>\>\>{\bf end if}\\
1\>0\>\>{\bf end for}\\
1\>1\>\>{\bf return} $(k,g[k])$ which maximizes
$\mu(k,g[k])$ for $1 \leq k \leq i_0$
\end{tabbing}
\hrule
\caption{Algorithm for finding maximum-density segment with width at least $L$}
\label{fig:algorithmL}
\end{figure}

\begin{thm}
\label{the:mainL}
Given a sequence $A$, the algorithm {\tt MaximumDensitySegmentL} finds
the maximum-density segment of those with width at least $L$.
\end{thm}
\begin{pf}
To prove the correctness, assume that $\hat{\mu}$ is the density of an
optimal such segment.  First, we note that for any value $i$,
Lemma~\ref{lem:matchL} assures that $g[i]$ is set such that $g[i] \geq
L_i$.  Therefore, $\mu(k,g[k]) \leq \hat{\mu}$ for all $k$ for which
$g[k]$ was defined.

We claim that for some $k$, value $g[k]$ is set such that $\mu(k,g[k])
\geq \hat{\mu}$.  Assume that the maximum density $\hat{\mu}$ is
achieved by some segment $A(i,i')$.  Since it must be that $L_i$ is
well-defined, we consider the pass of the loop starting at line~4 for
such an $i$.  Lemma~\ref{lem:matchL2} assures us that if {\tt
FindMatchL} is called, it either returns $i'$ or else it must be
the case that for some $j>i$, $g[j] < i'$ In this case, as $U$ is
unbounded, Lemma~\ref{lem:main} assures us that $\mu(j,g[j]) \geq
\mu(i,i') = \hat{\mu}$.  And thus {\tt MaximumDensitySegmentL} 
returns a segment with density $\hat{\mu}$.
\qed\end{pf}

\begin{thm}
\label{the:costL}
Given a sequence $A$ of length $n$, {\tt
MaximumDensitySegmentL} runs in $O(n)$ time.
\end{thm}
\begin{pf}
This is a direct consequence of Lemma~\ref{lem:costL}.
\qed\end{pf}

\subsection{Maximum-Density Segment with Width at Least $L$ and at Most $U$}
\label{sec:LU}

In this section, we consider the problem of finding a segment with the
maximum possible density among those of width at least $L$ and at most
$U$. At first glance, the sweeping of variable $u$ in the previous
algorithm appears similar to placing an explicit upper bound on the
width of the segments of interest for a given left index $i$.  In
locating the good partner for $i$, a sequential search is performed
over right-skew segments of $\rs{A(L_i+1,n)}$.  The repeated decision
of whether it is advantageous to include the bridge segment
$A(b,p[b])$ is determined in accordance with the bitonic property of
Lemma~\ref{lem:bitonic}.

The reason that this technique does not immediately apply to the case
with an explicit upper bound of $U$ is the following. If the right
endpoint of the bridge, $p[b]$, is strictly greater than $U_i$,
considering the effect of including the entire bridge may not be
relevant.  To properly apply
Lemmas~\ref{lem:atomic}--\ref{lem:bitonic}, we must consider segments
of $\rs{A(L_i+1,U_i)}$ as opposed to $\rs{A(L_i+1,n)}$.

\subsubsection{Another Sweep-Line Data Structure}
\label{sec:sweepU}

Recall that the structure of Section~\ref{sec:sweepL} focused on
finding segments beginning with $i$, ending in $[x,y]$ and subject to
a {\em lower} bound on the resulting segment width.  Therefore, as $i$
was decreased, the effective lower bound, $L_i$, on the matching
endpoint can only decrease. The decomposition of interest was
$\rs{A(L_i,y)}$, and such decompositions were simultaneously
represented for all possible values of $L_i$ by the right-skew
pointers, $p[k]$.

In this section, we develop another sweep-line data structure that we
used to locate segments beginning with $i$, ending in $[x,y]$ and
subject to an {\em upper} bound on the resulting segment width (but
with no explicit lower bound).  For a given $i$, the decomposition of
interest is $\rs{A(x+1,U_i)}$.  However, since $U_i$ decreases with
$i$, our new structure is based on representing the decreasingly
right-skew partitions for all {\em prefixes} of $A(x+1,y)$, rather
than all {\em suffixes}.  We assign values $q[k]$ for $x+1 \leq k
\leq y$ such that $A(q[k],k)$ is the {\em rightmost} segment of
$\rs{A(x+1,k)}$.  Though there are clear symmetries between this
section and Section~\ref{sec:sweepL}, there is not a perfect symmetry;
in fact the structure introduced in this section is considerably
simpler. The lack of perfect symmetry is because the concept of {\em
right}-skew segments, used in both sections, is oriented.

\begin{figure}[t]
\centerline{\psfig{figure=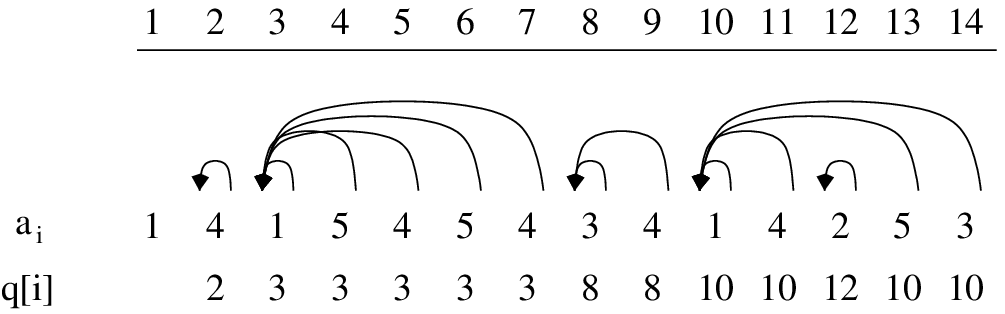,width=9cm}}
\caption{Example of data structure after {\tt InitializeU}$(1,14)$,
with $w_i = 1$ for all $i$.}
\label{fig:exampleU}
\end{figure}

The initialization routine for this new structure is presented in
Figure~\ref{fig:initializeU}.  An example of an initialized structure
is given in Figure~\ref{fig:exampleU}. 
\begin{figure}[t]
\hrule
\begin{tabbing}
x\=xxx\=xxx\=xxx\=xxx\=xxx\=xxx\=\kill
\\
\>\>{\bf procedure} {\tt InitializeU}$(x,y)$ 
\mbox{\hspace{0.75in}} {\em assumes $1 \leq x \leq y \leq n$}\\
\>1\>\>{\bf for} $i \leftarrow x+1$ {\bf to} $y$ {\bf do}\\
\>2\>\>\>$q[i] \leftarrow i$\\
\>3\>\>\>{\bf while} $\left((q[i]>x) \mbox{ and }
(\mu(q[q[i]-1],q[i]-1)\leq \mu(q[i],i))\right)$ {\bf do}\\ 
\>4\>\>\>\>$q[i] \leftarrow q[q[i]-1]$\\
\>5\>\>\>{\bf end while}\\
\>6\>\>{\bf end for}\\
\>7\>\>$u \leftarrow y$
\end{tabbing}
\hrule
\caption{{\tt InitializeU} operation.}
\label{fig:initializeU}
\end{figure}
The redesign of the initialization routine relies on a simple duality
when compared with the corresponding routine of
Section~\ref{sec:sweepL}.  One can easily verify that an execution of
this routine on a segment $A(x,y)$ sets the values of array $q$
precisely as the original version would set the values of array $p$ if
run on a reversed and negated copy of $A(x,y)$.  Based on this
relationship, we claim the following dual of
Lemma~\ref{lem:p} without further proof.


\begin{lem}
\label{lem:q}
Immediately after 
{\tt InitializeU}$(x,y)$, the segment
$A(q[k],k)$ is the rightmost segment of $\rs{A(x+1,k)}$, for all $k$
in the range $[x+1,y]$.
\end{lem}

We now present the main query routine, {\tt FindMatchU}, given in
Figure~\ref{fig:findmatchU}, and discuss its behavior.


\begin{figure}[t]
\hrule
\begin{tabbing}
x\=xxx\=xxx\=xxx\=xxx\=xxx\=xxx\=\kill
\\
\>\>{\bf procedure} {\tt FindMatchU}$(i)$\\
\>1\>\>{\bf while} $(u > U_i)$ {\bf do} \hspace{2.25in} // decrease $u$\\
\>2\>\>\>$u \leftarrow u-1$\\
\>3\>\>{\bf end while}\\
\>4\>\>{\bf while} $(u > x)$ and $(\mu(i,q[u]-1) > \mu(i,u))$ {\bf do}
\hspace{0.25in} // bitonic search\\
\>5\>\>\>$u \leftarrow  q[u]-1$\\
\>6\>\>{\bf end while}\\
\>7\>\>{\bf return} $u$
\end{tabbing}
\hrule
\caption{{\tt FindMatchU}$(i)$ operation.}
\label{fig:findmatchU}
\end{figure}

\begin{lem}
\label{lem:matchU}
Assume {\tt FindMatchU}$(i)$ is called with a value $i$ less than that
of all previous invocations and such that $x \leq U_i \leq y$.  Let
$m_0$ be the most recently returned value from {\tt FindMatchU}$()$ or
$y$ if this is the first such call. Let $A(i,m)$ be a maximum-density
segment of those starting with $i$, having width at most $U$, and
ending with $m \in [x,m_0]$.  Then {\tt FindMatchU}$(i)$ returns the
value $m$.
\end{lem}

\begin{pf}
Combining the constraints that $w(i,m) \leq U$ and that $m \in
[x,m_0]$, it must be that $m \leq \min(U_i,m_0)$.  When entering the
procedure, the variable $u$ has value $m_0$. The loop starting at
line~1 ensures that variable $u = \min(U_i,m_0)$ upon the loop's exit.
The discussion in Section~\ref{sec:skew} assures us that the optimal
$m \in [x,u]$ must either be $x$ or else among the right endpoints of
$\rs{A(x+1,u)}$.  Based on Lemma~\ref{lem:q}, $A(q[u],u)$ is the
rightmost segment of $\rs{A(x+1,u)}$ and so the loop condition at
line~4 of {\tt FindMatchU} is a direct application of
Lemma~\ref{lem:bitonic}.
\qed\end{pf}

\begin{lem}
\label{lem:costU}
The data structure supports its operations with amortized running
times of $O(y-x+1)$ for {\tt InitializeU}$(x,y)$, and $O(1)$ for {\tt
FindMatchU}$(i)$, so long as $U_i \geq x$ for all $i$.
\end{lem}
\begin{pf}
The initialization procedure has an $O(y-x+1)$-time worst-case bound,
as was the case for the similar routine in Section~\ref{sec:sweepL}.

To account for the cost of {\tt FindMatchU}, we note that $u$ is
initialized to value $y$ at line~7 of {\tt InitializeU}.  It is only
modified by lines~$2$~and~$5$ of the routine, and we claim that both
lines strictly decrease the value. This is obvious for line~2, and for
line~5, it follows for {\tt FindMatchU}, since $q[u] \leq u$ in
accordance with Lemma~\ref{lem:q}.  We also claim that $u$ is never
set less than $x$. Within the loop of lines~1--3, this is due to the
assumption that $U_i \geq x$.  For the loop of lines~4--7, it is true
because $q[u] \geq x+1$ in accordance with Lemma~\ref{lem:q}.
Therefore, these loops execute at most $O(y-x+1)$ times combined and
this cost can be amortized against the initialization cost. An $O(1)$
amortized cost per call can account for checking the initial test
condition before entering either loop.
\qed\end{pf}

\subsubsection{An $O(n)$-time Algorithm for the Uniform Model}
\label{sec:algorithmLU}

In this section, we present a linear-time algorithm for the {\em
uniform} maximum-density segment problem subject to both a lower bound
$L$ and an upper bound $U$ on the segment width, with $L < U$. 

Our strategy is as follows.  We pre-process the original sequence by
breaking it into blocks of cardinality exactly $U-L$ (except, possibly
for the last block).  For each such block, we maintain two sweep-line
data structures, one as in Section~\ref{sec:sweepL} and one as in
Section~\ref{sec:sweepU}.

For a given left index $i$, a valid good partner must lie in the range
$[L_i,U_i]$.  Because we consider the {\em uniform} model, such an
interval had cardinality precisely $(U-L+1)$ and thus overlaps
exactly two of the pre-processed blocks.  For $\alpha = (U-L) \lceil
L_i/(U-L) \rceil$, we search for a potential partner in the range
$[L_i,\alpha]$ using the data structure of Section~\ref{sec:sweepL},
and for a potential partner in the range $[\alpha+1,U_i]$ using the
data structure of Section~\ref{sec:sweepU}.  Though we are not assured
of finding the true good partner for each $i$, we again find the
global optimum, in accordance with Lemma~\ref{lem:main}.
Our complete algorithm is given in Figure~\ref{fig:algorithmLU}.

\begin{figure}[t]
\hrule
\begin{tabbing}
x\=xxx\=xxx\=xxx\=xxx\=xxx\=xxx\=\kill
\\
\>\>{\bf procedure} {\tt MaximumDensitySegmentLU}$(A,L,U)$\\
\>1\>\>[calculate values $L_i$ and $U_i$, as discussed
in Section~\ref{sec:notation}]\\
\>2\>\>$leftend \leftarrow 1$\\
\>3\>\>{\bf while} $(leftend < n)$ {\bf do}   \hspace{0.3in} // initialize blocks\\
\>4\>\>\>$rightend \leftarrow \min(n,leftend+(U-L)-1)$\\
\>5\>\>\>$z \leftarrow (leftend-1)/(U-L)$\\
\>6\>\>\> $\mbox{\tt Block}_L[z] \leftarrow \mbox{\tt
InitializeL}(leftend,rightend)$\\
\>7\>\>\> $\mbox{\tt Block}_U[z] \leftarrow \mbox{\tt
InitializeU}(leftend,rightend)$\\
\>8\>\>\> $leftend \leftarrow leftend + (U-L)$\\
\>9\>\>{\bf end while}\\
1\>0\>\>$i_0 \leftarrow$ maximum index such that $L_{i_0}$ is well-defined\\
1\>1\>\>{\bf for} $i \leftarrow i_0$ {\bf downto} $1$ {\bf do}\\
1\>2\>\>\>$z \leftarrow \lceil L_i/(U-L) \rceil$  \hspace{0.25in} // determine which blocks to search\\
1\>3\>\>\>$g_L[i] \leftarrow$ {\tt FindMatchL}$(i)$  for $\mbox{\tt Block}_L[z]$\\
1\>4\>\>\>$g_U[i] \leftarrow$ {\tt FindMatchU}$(i)$  for $\mbox{\tt Block}_U[z+1]$\\
1\>5\>\>\>{\bf if} $(\mu(i,g_L[i]) \geq \mu(i,g_U[i]))$ {\bf then}\\
1\>6\>\>\>\> $g[i] \leftarrow g_L[i]$\\
1\>7\>\>\>{\bf else}\\
1\>8\>\>\>\> $g[i] \leftarrow g_U[i]$\\
1\>9\>\>\>{\bf end if}\\
2\>0\>\>{\bf end for}\\
2\>1\>\>{\bf return} $(k,g[k])$ which maximizes
$\mu(k,g[k])$ for $1 \leq k \leq i_0$
\end{tabbing}
\hrule
\caption{Algorithm for finding maximum-density segment with width at
least $L$ and at most $U$}
\label{fig:algorithmLU}
\end{figure}

\begin{thm}
\label{the:mainLU}
Given a sequence $A$ of length $n$ for which $w_i=1$ for all $i$, and
parameters $L<U$, the algorithm {\tt MaximumDensitySegmentLU} finds
the maximum-density segment of those with width at least $L$ and at
most $U$.
\end{thm}
\begin{pf}
Because $w_i=1$ for all $i$, the interval $[L_i,U_i]$ has cardinality
precisely $(U-L+1)$. We note that $z$ is chosen at line~12 so that
$L_i$ lies within $\mbox{\tt Block}_L[z]$ and that $U_i$ lies within
$\mbox{\tt Block}_U[z+1]$.

To prove the correctness, assume that $\hat{\mu}$ is the density of an
optimal such segment.  First, we claim that $L \leq w(i,g[i]) \leq U$
for any $i$ and thus that the algorithm cannot possibly return a
density greater than $\hat{\mu}$.  This is a direct result of
Lemma~\ref{lem:matchL} in regard to the block searched at line~13 and
Lemma~\ref{lem:matchU} in regard to the block searched at line~14.

We then claim that for some $k$, value $g[k]$ is set such that
$\mu(k,g[k]) \geq \hat{\mu}$.  Assume that the maximum density
$\hat{\mu}$ is achieved by some segment $A(i,i')$.  Since it must be
that $L_i$ is well-defined, we consider the pass of the loop starting
at line~11 for such an $i$.  We show that either $g[i] = i'$ or
else there exists some $j>i$ such that $\mu(j,g[j]) \geq \mu(i,i')$.
If $i'$ lies within $\mbox{\tt Block}_L[z]$ then we apply the same
reasoning about the call to {\tt FindMatchL} at line~13, as we did in
the proof of Theorem~\ref{the:mainL}.  If $i'$ lies within $\mbox{\tt
Block}_U[z+1]$ then we consider the behavior of the call to {\tt
FindMatchU}$(i)$ at line~14.  If that call does not return $i'$ then
it must be that $i' > m_0$ as defined in Lemma~\ref{lem:matchU}.
Since the return values of this method are non-increasing, we let $j$
be the largest index for which the returned $g[j] < i'$.  At the onset
of that call to {\tt FindMatchU} it must have been that $m_0 \geq i'$
and therefore we can apply Lemma~\ref{lem:main} to deduce that
$\mu(j,g[j]) \geq \mu(i,i')$.  Thus {\tt MaximumDensitySegmentLU}
returns a segment with density $\hat{\mu}$.
\qed\end{pf}

\begin{thm}
\label{the:costLU}
Given a sequence $A$ of length $n$ for which $w_i=1$ for all $i$, the
algorithm {\tt MaximumDensitySegmentLU} runs in
$O(n)$ time.
\end{thm}
\begin{pf}
This is a consequence of Lemmas~\ref{lem:costL}~and~\ref{lem:costU}.
The calls to the initialization routines in the loop of lines~3--9
runs in $O(n)$ time, as each initialization routine is called for a
set of blocks which partition the original sequence.  Similarly, the
cost of all the calls to {\tt FindMatchL} and {\tt FindMatchU} in the
loop of lines~11--20 can be amortized against the corresponding
initializations.
\qed\end{pf}

\subsubsection{An $O(n + n \log(U-L+1))$-time Algorithm for a More
General Model}
\label{sec:algorithmLUgeneral}

With general values of $w_i$, the algorithm in
Section~\ref{sec:algorithmLU} is insufficient for one of two reasons.
If values of $w_i>1$ are allowed, it may be that an entire interval
$[L_i,U_i]$ falls in a single block, in which case neither of the
sweep-line data structures suffice.  Alternatively, if values of
$w_i<1$ are allowed, then an interval $[L_i,U_i]$ might span an
arbitrary number of blocks.  Always searching all such blocks might
result in $\omega(n)$ overall calls to {\tt FindMatchL} or {\tt
FindMatchU}.

In this section, we partially address the general case, providing an
$O(n + n\log (U-L+1))$-time algorithm when $w_i \geq 1$ for all $i$.
This condition assures us that the interval $[L_i,U_i]$ has
cardinality at most $(U-L+1)$.  Rather than rely on a single partition
of the original sequence into fixed-sized blocks, we will create
$O(\log (U-L+1))$ partitions, each of which uses fixed-sized blocks,
for varying sizes.  Then we show that the interval $[L_i,U_i]$ can be
covered with a collection of smaller blocks in which we can search.

For ease of notation, we assume, without loss of generality, that the
overall sequence $A$ is padded with values so that it has length $n$
which is a power of two.  We consider $n$ blocks of size $1$, $n/2$
blocks of size $2$, $n/4$ blocks of size $4$, and so on until
$n/2^\beta$ blocks of size $2^\beta$, where $\beta = \lfloor \log_2
(U-L+1)
\rfloor$. Specifically, we define block $B_{j,k} = A(1+j*2^k, (j+1)*2^k)$
for all $0 \leq k \leq \beta$ and $0 \leq j < n/2^k$. We begin
with the following lemma.

\begin{figure}[t]
\centerline{\psfig{figure=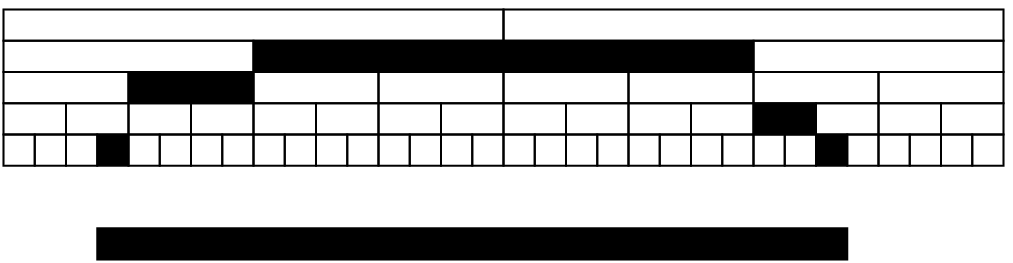,width=3.0in}}
\caption{A collection of blocks for a given interval}
\label{fig:blockexample}
\end{figure}

\begin{lem}
\label{lem:blocks}
For any interval $A(p,q)$ with cardinality at most $U-L+1$, we can
compute, in $O(1+\beta)$ time, a collection of $O(1+\beta)$ disjoint
blocks such that $A(p,q)$ equals the union of the blocks.
\end{lem}
The algorithm {\tt CollectBlocks} is given in
Figure~\ref{fig:blockalgo}, and a sample result is shown in
Figure~\ref{fig:blockexample}.  It is not difficult to 
verify the claim.

\begin{figure}[t]
\hrule
\begin{tabbing}
x\=xxx\=xxx\=xxx\=xxx\=xxx\=xxx\=\kill
\\
\>\>{\bf procedure} {\tt CollectBlocks}$(p,q)$\\
\>1\>\>$s \leftarrow p$; $k \leftarrow 0$;\\
\>2\>\>{\bf while} $(2^k+s-1 \leq q)$ {\bf do}\\
\>3\>\>\>{\bf while} $(2^{k+1}+s-1 \leq q)$ {\bf do}\\
\>4\>\>\>\>$k \leftarrow k+1$\\
\>5\>\>\>{\bf end while}\\
\>6\>\>\> $j \leftarrow \lceil s/2^k \rceil - 1$\\
\>7\>\>\> Add block $B_{j,k}$ to the collection\\
\>8\>\>\>$k \leftarrow k+1$\\
\>9\>\>{\bf end while}\\
1\>0\>\>{\bf while} $(s \leq q)$ {\bf do}\\
1\>1\>\>\>{\bf while} $(2^{k}+s-1 > q)$ {\bf do}\\
1\>2\>\>\>\>$k \leftarrow k-1$\\
1\>3\>\>\>{\bf end while}\\
1\>4\>\>\> $j \leftarrow \lceil s/2^k \rceil - 1$\\
1\>5\>\>\> Add block $B_{j,k}$ to the collection\\
1\>6\>\>\>$k \leftarrow k-1$\\
1\>7\>\>{\bf end while}\\
\end{tabbing}
\hrule
\caption{Algorithm for finding collection of blocks to cover an interval}
\label{fig:blockalgo}
\end{figure}

\begin{figure}[t]
\hrule
\begin{tabbing}
x\=xxx\=xxx\=xxx\=xxx\=xxx\=xxx\=\kill
\\
\>\>{\bf procedure} {\tt MaximumDensitySegmentLU2}$(A,L,U)$\\
\>1\>\>[calculate values $L_i$ and $U_i$, as discussed
in Section~\ref{sec:notation}]\\
\>2\>\>$\beta \leftarrow \lfloor \log_2 (U-L+1) \rfloor$; $\mu_{max}
\leftarrow - \infty$;\\
\>3\>\>{\bf for} $k \leftarrow 0$ {\bf to } $\beta$ {\bf do}\\
\>4\>\>\>{\bf for} $j \leftarrow 0$ {\bf to } $n/2^k - 1$ {\bf do}\\
\>5\>\>\>\>$B_{j,k} \leftarrow$ {\tt InitializeL}$(1+j*2^k, (j+1)*2^k)$\\
\>6\>\>\>{\bf end for}\\
\>7\>\>{\bf end for}\\
\>8\>\>$i_0 \leftarrow$ maximum index such that $L_i$ is defined\\
\>9\>\>\>{\bf foreach} $B_{j,k}$ in {\tt CollectBlocks}$(L_i,U_i)$ {\bf do}\\
1\>0\>\>\>\>$temp \leftarrow $ {\tt FindMatchL}$(i)$\\
1\>1\>\>\>\> {\bf if} $\mu(i,temp) > \mu_{max}$ {\bf then}\\
1\>2\>\>\>\>\>$\mu_{max} \leftarrow  \mu(i,temp)$; record endpoints $(i,temp)$\\
1\>3\>\>\>\>{\bf end if}\\
1\>4\>\>\>{\bf end foreach}\\
1\>5\>\>{\bf end for}\\
\end{tabbing}
\hrule
\caption{Algorithm for maximum-density segment with width at
least $L$, at most $U$}
\label{fig:algorithmLUgeneral}
\end{figure}

\begin{thm}
\label{the:mainLUgeneral}
Given a sequence $A$ of length $n$ for which $w_i \geq 1$ for all $i$,
a maximum-density segment of those with width at least $L$ and at most
$U$ can be found in $O(n+n\log(U-L+1))$ time.
\end{thm}
\begin{pf}
The algorithm is given in Figure~\ref{fig:algorithmLUgeneral}.  First,
we discuss the correctness.  Assume that the global maximum is
achieved by $A(i,i')$.  We must simply show that this pair, or one with
equal density, was considered at line~14.  By Lemma~\ref{lem:blocks},
$i'$ must lie in some $B_{j,k}$ returned by {\tt CollectBlocks}$(L_i,U_i)$.
Because $\mu(i,i')$ is a global maximum, the combination of
Lemmas~\ref{lem:main}~and~\ref{lem:matchL} assures us that {\tt
FindMatchL()} at line~10 will return $i'$ or else some earlier pair
which was found has density at least as great.

We conclude by showing that the running time is $O(n+n\beta)$.  Notice
that for a fixed $k$, blocks $B_{j,k}$ for $0 \leq j < n/2^k - 1$
comprise a partition of the original input $A$, and thus the sum of
their cardinalities is $O(n)$.  Thefefore, for a fixed $k$, lines~4--6
run in $O(n)$ time by Lemma~\ref{lem:costL}, and overall lines~3--7
run in $O(n\beta)$ time.  Each call to {\tt CollectBlocks} from line~9
runs in $O(1+\beta)$ time by Lemma~\ref{lem:blocks}, and produces
$O(1+\beta)$ blocks.  Therefore, the body of that loop, lines~9--14,
executes $O(n+n\beta)$ times over the course of the entire algorithm.

Finally, we must account for the time spent in all calls to {\tt
FindMatchL} from line~10.  Rather than analyze these costs
chronologically, we account for these calls by considering each block
$B_{j,k}$ over the course of the algorithm.  By Lemma~\ref{lem:costL},
each of these calls has an amortized cost of $O(1)$, where that cost
is amortized over the initialization cost for that block.
\qed\end{pf}

\section*{Acknowledgments}
We wish to thank Yaw-Ling Lin for helpful discussions.

\bibliographystyle{alpha}
\bibliography{all}

\newcommand{\etalchar}[1]{$^{#1}$}
\begin{thebibliography}{WLOG02}

\bibitem[AS98]{Alexandrov:1998:SSU}
N.~N. Alexandrov and V.~V. Solovyev.
\newblock Statistical significance of ungapped sequence alignments.
\newblock In {\em Proceedings of Pacific Symposium on Biocomputing}, volume~3,
  pages 461--470, 1998.

\bibitem[Bar00]{Barhardi:2000:IEG}
G.~Barhardi.
\newblock Isochores and the evolutionary genomics of vertebrates.
\newblock {\em Gene}, 241:3--17, 2000.

\bibitem[BB86]{Bernardi:1986:CCG}
G.~Bernardi and G.~Bernardi.
\newblock Compositional constraints and genome evolution.
\newblock {\em Journal of Molecular Evolution}, 24:1--11, 1986.

\bibitem[Ben86]{Bentley:1986:PP}
Jon~Louis Bentley.
\newblock {\em Programming Pearls}.
\newblock Addison-Wesley, Reading, MA, 1986.

\bibitem[Cha94]{Charlesworth:1994:GRP}
B.~Charlesworth.
\newblock Genetic recombination: patterns in the genome.
\newblock {\em Current Biology}, 4:182--184, 1994.

\bibitem[DMG95]{Duret:1995:SAV}
L.~Duret, D.~Mouchiroud, and C.~Gautier.
\newblock Statistical analysis of vertebrate sequences reveals that long genes
  are scarce in {GC}-rich isochores.
\newblock {\em Journal of Molecular Evolution}, 40:308--371, 1995.

\bibitem[EW92]{Eyre-Walker:1992:EBG}
Adam Eyre-Walker.
\newblock Evidence that both {G+C} rich and {G+C} poor isochores are replicated
  early and late in the cell cycle.
\newblock {\em Nucleic Acids Research}, 20:1497--1501, 1992.

\bibitem[EW93]{Eyre-Walker:1993:RMG}
Adam Eyre-Walker.
\newblock Recombination and mammalian genome evolution.
\newblock {\em Proceedings of the Royal Society of London Series B, Biological
  Science}, 252:237--243, 1993.

\bibitem[FCC01]{Fullerton:2001:LRR}
Stephanie~M. Fullerton, Antonio~Bernardo Carvalho, and Andrew~G. Clark.
\newblock Local rates of recombination are positively corelated with {GC}
  content in the human genome.
\newblock {\em Molecular Biology and Evolution}, 18(6):1139--1142, 2001.

\bibitem[Fil87]{Filipski:1987:CBM}
J.~Filipski.
\newblock Correlation between molecular clock ticking, codon usage fidelity of
  {DNA} repair, chromosome banding and chromatin compactness in germline cells.
\newblock {\em FEBS Letters}, 217:184--186, 1987.

\bibitem[FO99]{Francino:1999:IRM}
M.~P. Francino and H.~Ochman.
\newblock Isochores result from mutation not selection.
\newblock {\em Nature}, 400:30--31, 1999.

\bibitem[FS90]{Fields:1990:GPT}
C.~A. Fields and C.~A. Soderlund.
\newblock gm: a practical tool for automating {DNA} sequence analysis.
\newblock {\em Computer Applications in the Biosciences}, 6:263--270, 1990.

\bibitem[GGA{\etalchar{+}}98]{Guldberg:1998:DMG}
P.~Guldberg, K.~Gronbak, A.~Aggerholm, A.~Platz, P.~{thor Straten},
  V.~Ahrenkiel, P.~Hokland, and J.~Zeuthen.
\newblock Detection of mutations in {GC}-rich {DNA} by bisulphite denaturing
  gradient gel electrophoresis.
\newblock {\em Nucleic Acids Research}, 26(6):1548--1549, 1998.

\bibitem[HDV{\etalchar{+}}91]{Hardison:1991:SCA}
R.~C. Hardison, D.~Drane, C.~Vandenbergh, J.-F.~F. Cheng, J.~Mansverger,
  J.~Taddie, S.~Schwartz, X.~Huang, and W.~Miller.
\newblock Sequence and comparative analysis of the rabbit alpha-like globin
  gene cluster reveals a rapid mode of evolution in a {G+C} rich region of
  mammalian genomes.
\newblock {\em Journal of Molecular Biology}, 222:233--249, 1991.

\bibitem[HHJ{\etalchar{+}}97]{Henke:1997:BIP}
W.~Henke, K.~Herdel, K.~Jung, D.~Schnorr, and S.~A. Loening.
\newblock Betaine improves the {PCR} amplification of {GC}-rich {DNA}
  sequences.
\newblock {\em Nucleic Acids Research}, 25(19):3957--3958, 1997.

\bibitem[Hol92]{Holmquist:1992:CBT}
G.~P. Holmquist.
\newblock Chromosome bands, their chromatin flavors, and their functional
  features.
\newblock {\em American Journal of Human Genetics}, 51:17--37, 1992.

\bibitem[Hua94]{Huang:1994:AIR}
X.~Huang.
\newblock An algorithm for identifying regions of a {DNA} sequence that satisfy
  a content requirement.
\newblock {\em Computer Applications in the Biosciences}, 10(3):219--225, 1994.

\bibitem[IAY{\etalchar{+}}96]{Ikehara:1996:PON}
K.~Ikehara, F.~Amada, S.~Yoshida, Y.~Mikata, and A.~Tanaka.
\newblock A possible origin of newly-born bacterial genes: significance of
  {GC}-rich nonstop frame on antisense strand.
\newblock {\em Nucleic Acids Research}, 24(21):4249--4255, 1996.

\bibitem[Inm66]{Inman:1966:DMP}
R.~B. Inman.
\newblock A denaturation map of the 1 phage {DNA} molecule determined by
  electron microscopy.
\newblock {\em Journal of Molecular Biology}, 18:464--476, 1966.

\bibitem[JFN97]{Jin:1997:WIN}
Ruzhong Jin, Maria-Elena {Fernandez-Beros}, and Richard~P. Novick.
\newblock Why is the initiation nick site of an {AT}-rich rolling circle
  plasmid at the tip of a {GC}-rich cruciform?
\newblock {\em The EMBO Journal}, 16(14):4456--4466, 1997.

\bibitem[LJC02]{Lin:2002:EAL}
Y.~L. Lin, T.~Jiang, and K.~M. Chao.
\newblock Algorithms for locating the length-constrained heaviest segments,
  with applications to biomolecular sequence analysis.
\newblock {\em Journal of Computer and System Sciences}, 2002.
\newblock To appear.

\bibitem[MHL01]{Murata:2001:TAF}
Shin-{ichi} Murata, Petr Herman, and Joseph~R. Lakowicz.
\newblock Texture analysis of fluorescence lifetime images of {AT}- and
  {GC}-rich regions in nuclei.
\newblock {\em Journal of Hystochemistry and Cytochemistry}, 49:1443--1452,
  2001.

\bibitem[MRO97]{Madsen:1997:ICE}
Cort~S. Madsen, Christopher~P. Regan, and Gary~K. Owens.
\newblock Interaction of {CArG} elements and a {GC}-rich repressor element in
  transcriptional regulation of the smooth muscle myosin heavy chain gene in
  vascular smooth muscle cells.
\newblock {\em Journal of Biological Chemistry}, 272(47):29842--29851, 1997.

\bibitem[MTB76]{Macaya:1976:AOE}
G.~Macaya, J.-P. Thiery, and G.~Bernardi.
\newblock An approach to the organization of eukaryotic genomes at a
  macromolecular level.
\newblock {\em Journal of Molecular Biology}, 108:237--254, 1976.

\bibitem[NL00]{Nekrutenko:2000:ACH}
Anton Nekrutenko and Wen-Hsiung Li.
\newblock Assessment of compositional heterogeneity within and between
  eukaryotic genomes.
\newblock {\em Genome Research}, 10:1986--1995, 2000.

\bibitem[RLB00]{Rice:2000:EEM}
P.~Rice, I.~Longden, and A.~Bleasby.
\newblock {EMBOSS}: The {E}uropean molecular biology open software suite.
\newblock {\em Trends in Genetics}, 16(6):276--277, June 2000.

\bibitem[SA93]{Scotto:1993:GRD}
L.~Scotto and R.~K. Assoian.
\newblock A {GC}-rich domain with bifunctional effects on {mRNA} and protein
  levels: implications for control of transforming growth factor beta 1
  expression.
\newblock {\em Molecular and Cellular Biology}, 13(6):3588--3597, 1993.

\bibitem[SAL{\etalchar{+}}95]{Sharp:1995:DSE}
P.~M. Sharp, M.~Averof, A.~T. Lloyd, G.~Matassi, and J.~F. Peden.
\newblock {DNA} sequence evolution: the sounds of silence.
\newblock {\em Philosophical Transactions of the Royal Society of London Series
  B, Biological Sciences}, 349:241--247, 1995.

\bibitem[Sel84]{Sellers:1984:PRG}
Peter~H. Sellers.
\newblock Pattern recognition in genetic sequences by mismatch density.
\newblock {\em Bulletin of Mathematical Biology}, 46(4):501--514, 1984.

\bibitem[SFR{\etalchar{+}}99]{Stojanovic:1999:CFM}
N.~Stojanovic, L.~Florea, C.~Riemer, D.~Gumucio, J.~Slightom, M.~Goodman,
  W.~Miller, and R.~Hardison.
\newblock Comparison of five methods for finding conserved sequences in
  multiple alignments of gene regulatory regions.
\newblock {\em Nucleic Acids Research}, 27:3899--3910, 1999.

\bibitem[SMRB83]{Soriano:1983:DIR}
P.~Soriano, M.~Meunier-Rotival, and G.~Bernardi.
\newblock The distribution of interspersed repeats is nonuniform and conserved
  in the mouse and human genomes.
\newblock {\em Proceedings of the National Academy of Sciences of the United
  States of America}, 80:1816--1820, 1983.

\bibitem[Sue88]{Sueoka:1988:DMP}
N.~Sueoka.
\newblock Directional mutation pressure and neutral molecular evolution.
\newblock {\em Proceedings of the National Academy of Sciences of the United
  States of America}, 80:1816--1820, 1988.

\bibitem[WLOG02]{Wang:2002:RFP}
Zhijie Wang, Eli Lazarov, Mike O'Donnel, and Myron~F. Goodman.
\newblock Resolving a fidelity paradox: {Why} {Escherichia} coli {DNA}
  polymerase {II} makes more base substitution errors in at- compared to
  {GC}-rich {DNA}.
\newblock {\em Journal of Biological Chemistry}, 2002.
\newblock To appear.

\bibitem[WSE{\etalchar{+}}99]{Wu:1999:IMD}
Y.~Wu, R.~P. Stulp, P.~Elfferich, J.~Osinga, C.~H. Buys, and R.~M. Hofstra.
\newblock Improved mutation detection in {GC}-rich {DNA} fragments by combined
  {DGGE} and {CDGE}.
\newblock {\em Nucleic Acids Research}, 27(15):e9, 1999.

\bibitem[WSL89]{Wolfe:1989:MRD}
K.~H. Wolfe, P.~M. Sharp, and Wen-Hsiung Li.
\newblock Mutation rates differ among regions of the mammalian genome.
\newblock {\em Nature}, 337:283--285, 1989.

\bibitem[ZCB96]{Zoubak:1996:GDH}
S.~Zoubak, O.~Clay, and G.~Bernardi.
\newblock The gene distribution of the human genome.
\newblock {\em Gene}, 174:95--102, 1996.

\end{thebibliography}

\end{document}